# Low-magnitude seismic activity between the Kamchatka July 20 and July 29, 2025, earthquakes. Spatio-temporal evolution recovered using waveform cross-correlation

Ivan O. Kitov

**Abstract**

The M8.8 Kamchatka earthquake on July 29, 2025 was one of the largest in the first quarter of the 21st century. It deserves a thorough analysis including the preparation process. A smaller M7.4 earthquake occurred on July 20 with its epicenter within the confidence ellipse for the July 29 event. The aftershock sequence of the July 20 earthquake and the evolution of seismicity within the Kamchatka Peninsula region during 10 days period before the July 29 event may provide important information on the earthquake preparation and initiation processes. The CTBTO's International Monitoring System is one of the most sensitive global seismic networks comprising high-resolution array stations with enhanced sensitivity relative to three-component stations at the same locations. The International Data Centre of the CTBTO processes IMS data automatically and interactively to create a Reviewed Event Bulletin (REB), which serves as a source of information for the International Seismological Centre. Waveform cross-correlation (WCC) allows for additional detection capabilities to the IMS data and IDC processing when repeated seismicity is analyzed. The aftershock sequence of the July 20 earthquake is recovered using the WCC-based detection and phase association techniques as applied to the IMS data in order to accurately describe the spatio-temporal evolution of seismic process just before the July 29 event. With the reduced detection threshold, smaller events are found in the zones where the REB has no located sources. This finding opens up the possibility for a more detailed study of seismic and mechanical processes before the July 29 mainshock.

**Introduction**

The M8.8 earthquake on July 29, 2025 (J29) near the Kamchatka Peninsula is one of the largest in this century. Its geophysical features have to be thoroughly studied in order to understand the pre-seismic, co-seismic, and post-seismic processes in relation to the global and regional tectonic activity. One of the routine but important observations is the M7.4 earthquake that occurred on July 20 (J20), in the immediate vicinity of the hypocenter of the J29 event. The characteristics of the J20 earthquake are able to put important constraints on the physical and mechanical state of the Kamchatka seismic zone just before the biggest event was initiated. This earlier event could be a foreshock or an independent earthquake in the long-term seismicity evolution generated by the subduction of the Pacific plate.

The immediate and observable effect of the M7.4 earthquake on the stress field in the Kamchatka region is likely to be confined to the area where its aftershocks occur. This post-seismic process can be measured to assist in the estimation of its impact on the surrounding area. This made it important to improve the detection and location of the J20 aftershocks as well as routine seismicity in the broader region where the J29 impact was significant. The waveform cross-correlation (WCC) technique is relatively modern and is used to reduce the detection threshold as well as to improve the accuracy of location and magnitude estimates of repeating seismic sources [Israelsson, 1990; Joswig, 1990; Schaff, Richards, 2004, 2014; Kvaerna *et al*., 2010; Bobrov *et al*., 2014]. The gain from WCC-based methods of detection, parameter estimation, and phase association can reach an order of magnitude or more [Schaff, Richards, 2004, 2014; Schaff, Waldhauser, 2010].

Cross-correlation is a core processing routine used in the matched filter, which is the optimal linear filter [Turin, 1960] maximizing the signal-to-noise ratio (*SNR*) in perfect conditions with additive stochastic noise. In practice, microseismic noise is not stochastic and the matched filter is a sub-optimal technique in real-world applications. However, detection gain from WCC is substantial for seismic studies because the earthquake recurrence curve implies an approximately tenfold increase in the number of events with the magnitude decreasing by one unit. Much higher accuracy of location and magnitude estimation is provided by the WCC-based methods for the newly found events [Schaff, Richards, 2004, 2014; Gibbons, Ringdal, 2004; 2006; Gibbons *et al*., 2017; Waldhauser, Schaff, 2008; Schaff, Waldhauser, 2010; Selby, 2010; Bobrov *et al*., 2014; Kitov *et al*., 2025]. These methods are important for estimating the stress field evolution and geophysical medium structure along faults and in the surrounding volume.

Our main objective is related to the recovering the aftershock sequence of the J20 earthquake in the 10-day period prior to the J29 earthquake. Using seismic data from the International Monitoring System (IMS), the International Data Centre (IDC) of the CTBTO, has produced a Reviewed Event Bulletin (REB), which includes 640 events between 06:02:48 on 20 July and 23:24:48 29 July within a rectangular area bounded by coordinates 45°N-65°N, 145°E-165°E. For the period between January 7, 2001, and December 4, 2025, the IDC reported a total of 26646 events in this area, with 5987 of them occurring after the J29 mainshock.

WCC-based processing allows for finding additional 50% to 100% events with the parameters matching the REB quality requirements, as shown in several studies using the IDC's interactive review of automatic cross-correlation bulletins [Bobrov et al., 2014; Kitov, Sanina, 2025]. Therefore, events that were missed by the REB can be recovered and their parameters can be added to the REB for more accurate estimates of post-J20 seismic activity.

The secondary objective is related to the REB rules for depth estimation. Events with large uncertainties above the 10 km depth limit are fixed to the free surface. To obtain a reliable depth estimate, at least three pP and sP phases have to be associated with each REB event [Coyne et al., 2012]. This procedure is specific to the IDC processing and it introduces not only a significant bias into the depth distribution of REB events, but also results in large shifts in epicenters as the location algorithm loses one degree of freedom. The WCC-based location methods attribute the new found events to the depths of the master events (ME), with the waveform templates at the associated stations used in cross-correlation with continuous data.

It is supposed that any ME is the best at finding new events close in location to its hypocenter. The level of cross-correlation decreases with increasing distance between the ME and sought events [Arrowsmith *et al*., 2006; Baish *et al*., 2008]. Therefore, the secondary objective is to relocate the REB events into their positions in the cross-correlation bulletin (XSEL) and use them together with the new XSEL events for analysis of the spatio-temporal distribution of the J20 aftershocks. As the J29 and its aftershocks is the ultimate target of the extended research, we have processed with the WCC-based methods the whole area of interest, where the seismicity may relate to the J20 and J29 events and their aftershocks.

**Data and method**

The IMS includes two types of seismic stations: arrays and 3-component (3-C). A seismic array is similar to a well-known directional antenna, but the seismic waves are characterized by changing velocity in the train of successive seismic phases. It makes a seismic array to work as a velocity filter. It can separate and identify regular seismic waves with different apparent velocities and propagation azimuths. It also has an advantage over a collocated 3-C station in terms of detection threshold, with the gain proportional to the square root of the number of sensors [Schweitzer at al., 2012]. As a result, array stations of the IMS have much larger input to the REB than 3-C stations.

Currently, there are 28 primary arrays on the IMS network and 6 auxiliary arrays [Coyne et al., 2012]. For the IDC, the principal difference between the primary and auxiliary networks is the input to event statistical significance. Auxiliary stations, which number will be 120 in the final IMS network, do not contribute directly to event statistics, but are used to improve the estimates of key event parameters such as the origin time, hypocenter location, and body and surface wave magnitude. We are not limited by the IDC's Event Definition Criteria (EDC) based on the statistical significance of the event hypotheses. All stations contribute to event statistical significance proportionally to their participation rates in the REB events in the J29 area from 2023. A new regional primary array station PDYAR was added to the IMS in the middle of 2022. It has a crucial input to the REB and XSEL.

The J20 and J29 events were detected by almost all the primary and auxiliary arrays. All available data was used, with decreasing station inputs depending on the distance and type of station. A list of the 50 IMS stations that are most sensitive to the earthquakes in the Kamchatka region from the beginning of 2023 is presented in Appendix 1, together with the bulletins of J20 and J29 borrowed from REB with only first P phases selected. The *SNR* values in these bulletins serve as indicators of relative station sensitivity for the events of interest.

Figure 1 shows the REB events for the 10-day period between J20 and J29. It includes aftershocks within a radius of ~200 km around the J20 epicenter and several earthquakes that were far from the mainshock to be classified as aftershocks. Almost all of the 640 REB events (629 aftershocks, the mainshock, and 10 events outside the aftershock area) were fixed at the free surface. Only 18 of them had reliable depth estimates. This indicates that the REB aftershocks are not the best MEs as their epicenter and depth estimates are biased. Figure 1 also presents the set of 100 MEs selected in a thorough pair-wise cross-correlation exercise of all REB events within the area [Kitov *et al.*, 2026]. They all have reliable depth and epicenter estimates, and their bulletins are listed in Appendix 2. The spatial distribution of these MEs corresponds to the subducting Pacific plate.

There are ten earthquakes in Figure 1 out of the aftershock area. Six of those ten have depth estimates. The overall distribution of the aftershocks/earthquakes between July 20 and 29 reveals a possible wide gap, which corresponds to the area of the J29 aftershocks, as marked by the events occurring 15 days following the J29 in Figure 1. The aftershock area of the J20 event is small compared to the whole region and even to the J29 aftershock area. The effect of the J20 earthquake likely had a limited, but not negligible, impact on the stress state of the subduction zone. The J20 was almost collocated with the J29, but didn't become it despite the short period of 10 days between them. The J29 was likely ready to start, but the J20 rupture had failed to break through the quiet zone (QZ) in Figure 1.

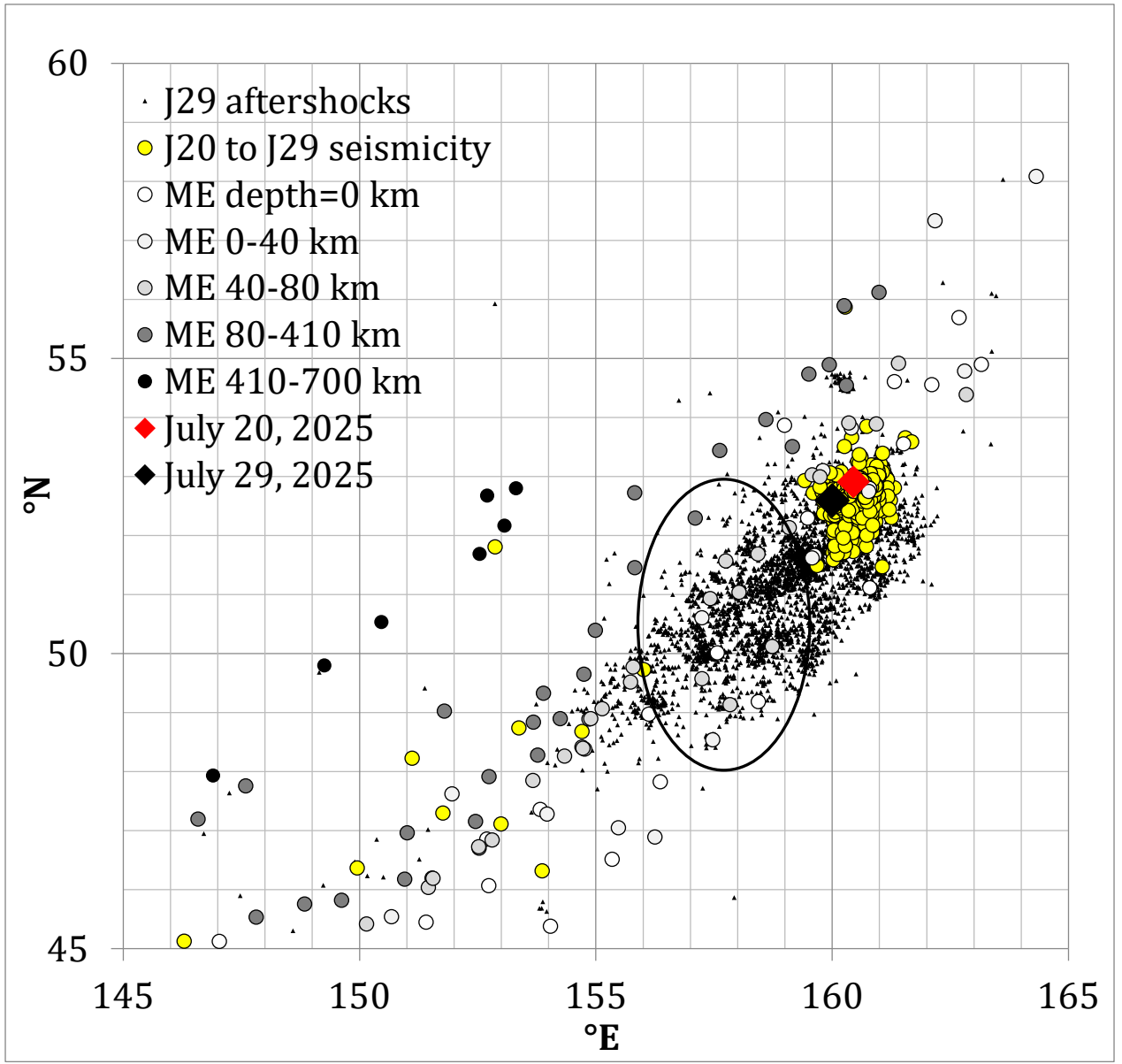

Figure 1. Aftershocks of J20 (yellow circles) and J29 (black triangles). Epicenters of events are as per the REB. Master events (ME) are distributed over the area of the J29 aftershocks and along the subducting plate. The quiet zone (no events during 10 days) from J20 to J29 has dimensions ~300x250 km and is marked by the oval. There are 12 shallow MEs within the quiet zone and a few very close to its border.

A remarkable feature in Figure 1 is the high concentration of the J29 aftershocks at the northeast border of the QZ (approximately, 48°N-53°N, 156°E-160°E) likely indicating a high energy release from the asperity destroyed by the J29 rupture. There were several earthquakes during the studied 10-day period downstream of the J29 rupture line, but beyond the QZ. The J29 aftershocks within and beyond the QZ of the J20 may not be fully related to the release of the preexisting local stresses, but also to the elastic energy released in the asperity destruction and transferred along the fault. It is also of importance that the J29 epicenter is downstream of the J20 one along its rupture line.

Each of the 100 MEs in Appendix 2 has many first P-wave arrivals at the associated stations. The effectiveness of the associated P-phases used as waveform templates can be characterized by their specificity, which is defined by the product of the spectral range above noise and the signal length. The specificity of a template is a key parameter for the rates of valid and false detections in a given noise/sought signals realization of actual data. The IDC standard *SNR* value determined for a given signal associated with an ME is related to the specificity of the corresponding template. Statistically, based on long-term seismological observations of natural sources, the higher is the *SNR* value for a given station associated with an ME, the broader is the spectral range above the ambient noise and the longer the signal remains above the noise.

The signals we search for are very close to the noise level. WCC is used specifically used to find the weakest possible signals, and one technique to increase *SNR* of weak signals is to filter them in various frequency bands and find the band where the *SNR* is the largest. The quality of the detected signals partially defines the statistical significance of the created event hypotheses, so the WCC detection procedure deserves detailed discussion to prove the existence and accurate location and magnitude estimation of the XSEL events. Since interactive analysis of the XSEL is not available, we rely on the REB as the reference to match with and potentially improve it using the XSEL event hypotheses.

The station performance and the process of detection threshold tuning can be illustrated by the properties of the *SNR* distribution. The closest to the J20 epicenter IMS array PETK (53.108°N, 157.699°E; 9 vertical sensors) is characterized by *SNR* figures measured by the IDC from ~15 to ~10,000 for distance range between 0.34° and 11° from the 640 events between July 20 and 29. Shallow earthquakes within this distances generate regular seismic phases $P_g$ and $P_n$ and a number of secondary phases, making the entire wave-train 30-180 seconds long. As an alternative example, array PDAR (38.430°N, 118.304°W; 13 vertical sensors) is characterized by *SNR* from ~5 to ~900 for signals from the J20 and its aftershocks in the range between ~55° and ~69°. At teleseismic distances, the P-wave signal duration is between 5 and 30 seconds depending on the magnitude and focal depth.

Waveform templates are used to calculate the continuous cross-correlation coefficient (*CC*) of time series (CC-trace). For a sought signal, we don't know the frequency band and the cross-correlation window length (CCWL) to obtain the maximum signal-to-noise ratio (*SNRcc*) estimated at the CC-traces. These parameters depend on the ambient noise properties and the source mechanism responsible for the signal amplitude in the direction towards the station. Potential frequency band for the first P-wave between 0.75 Hz and 6 Hz is covered by a comb of 5 band-pass filters. Parameter CCWL has to include the range where signal in the template is above ambient noise. For regional seismic phases at PETK, the length varies between 20 s and 120 s depending on station-event distance and depth. For teleseismic distances, CCWL depends on the frequency band and is between 4 s (high-frequency signals from low magnitude deep events) to 30 s (shallow events). For a given time count in the CC-trace, the final *SNRcc* is selected as the largest among the entire set of filters and CCWL values. The *SNRcc* detection threshold can also be filter/CCWL dependent.

In order to introduce a slightly complicated detection process using the CC-traces we first need to describe a standard energy detector. For standard *SNR* estimates obtained from real filtered data, the arrival time is the first count with *SNR* above the threshold. When the detection time is fixed and the corresponding long-term average (*LTA*- the *SNR* denominator) amplitude is "frozen" at its value at the time of detection, the peak *SNR* value is sought for in a predefined search time window (varying between seismic agencies from 5.5 to 15 seconds) to characterize the signal amplitude and period for body wave magnitude estimation. This peak in *SNR* is supposed to be relatively close to the arrival time. Increasing detection threshold affects the *SNR* peak as the *LTA* may include larger and larger portions of the signal energy together with the pre-signal noise. In some cases, detection with a relatively high peak value *SNR0* may disappear with the increasing threshold even if this threshold is below *SNR0*.

When all individual CC-traces are calculated at a 3-C station or array, they are averaged across all channels for each time count in order to obtain an averaged *CC* time series that is then used for detection purposes.

The short-term-average (*STA*) and *LTA* (absolute) amplitude, as well as their running ratio that is used as *SNRcc*, are first calculated for the averaged *CC*. The lengths of *STA* and *LTA* windows were estimated in our previous studies [Bobrov et al., 2014] and are fixed at 0.6 s and 60 s, respectively. As the maximum *SNRcc* values are selected from all filters and the CCWL value for each time count, successive values may belong to different sets of filters and CCWLs. A valid signal is detected when the *SNRcc* reaches a predetermined threshold. The difference with standard energy detector is that the arrival time of a WCC-detected signal is estimated near the peak of *SNRcc* time series, rather than at the first detection point. This *SNRcc* peak is sought in the cross-correlation window and for the filter corresponding to the time count where the *SNRcc* threshold was reached first. Therefore, the arrival time can be found tens of seconds after the detection threshold was first reached. The peak *SNRcc* has to be near the point in time where the template and the sought signal are synchronized the best. For high-amplitude templates and sought signals, the peak *CC* would be the best arrival time, as observed in the signals from the underground tests announced by the DPRK [Selby, 2010; Bobrov et al., 2014; Kitov et al., 2025]. For low-amplitude sought signals, the peak *CC* may be estimated less precisely due to the influence of noise. There may also be multiple *CC*-peaks with similar properties.

For the matched filter used as a detector, *SNRcc* is a good parameter to estimate the arrival time of the signal. However, it is also possible to allow the arrival time to correspond to peaks in the *CC* time series (using the same filter and CCWL as for detection) in a 1-second window around the *SNRcc* peak. The next *CC* detection has to be beyond the CCWL of the previous one. When a current detection hypothesis is rejected during a quality check procedure, the search for the next detection is started in several seconds from the rejected one. This procedure allows to reduce the effect of false detections related to spikes in data and side sensitivity of the WCC method to strong sources.

The frequency distribution of the *SNRcc* values over a given time period is the basis for estimating the detection threshold. If the threshold is set to a very high value, say *SNRcc*=100, the corresponding *SNRcc* distributions will not contain any valid detections and there will be no "frozen" *LTA* values skewing the time series of *SNRcc* estimates. For CC-traces, the threshold effect on the *LTA* estimates is much stronger as it may start to increase together with *CC* when the tail of the template first reaches the sought signal. The template always has a component coherent to the sought signal, even in the tail, and the *CC* starts to grow slowly towards its peak value. If the *LTA* is not "frozen". This effect suppresses the *SNRcc* growth far before it reaches the time of physical signal where the *SNRcc* value has to peak. The larger the *SNRcc* threshold, the lower the *SNRcc* peak value. The length of the CC-window is up to 120 s for the regional phases $P_g$ and $P_n$ followed by secondary phases $S_n$, $L_g$, and $R_g$. Therefore, the increase in the detection threshold in WCC processing has to be well coordinated with the CC-window length and the coherent properties of the templates and sought signals. For example, deep earthquakes generate impulsive signals and WCC processing is more or less similar to standard as the signals are impulsive and the peak *SNRcc* is close to the first count with *SNRcc* above the threshold. Shallower events may create very long wave-trains containing valuable information on the source and thus the whole length has to be used in the WCC process. This can suppress the *SNRcc* peak even at relatively low detection thresholds.

Figure 2 shows the *SNRcc* probability density functions (pdf) for the first five MEs from Appendix 2 for the array station PDAR on July 27, 2025, with bin width of 0.1. The *SNRcc* threshold was set at 100. All five pdf's are very similar, peaking around *SNRcc*=1.5 and then rolling off along a smooth curve to a heavy tails near *SNRcc*=3.0. This tail demonstrates the deviation from the smooth power law distribution for random values of *SNRcc* and indicates the presence of the sought signals with *SNRcc*>3 related to cross-correlation with corresponding templates. These signals have to be present in the data at PDAR because they are definitely generated by the J20 aftershocks. As the sought signals are characterized by duration of at least a few seconds, the number of *SNRcc* values above those predicted by a random distribution causes the observed deviation to occur.

In Figure 3, the case of the closest to the epicenter of the J20 earthquake IMS seismic array PETK is presented. There are six curves obtained for two MEs on three different dates: July 18 (2025199), July 20 (2025201), and July 30 (2025211), 2025. The day 18 July is selected as an example of a relatively quiet seismicity before the J20 event with hundreds of earthquakes in the REB. On 30 July 2025, there was a record number of daily events in the REB - more than 800 in the studied area. Station PETK is a small-aperture array and has to be most efficient at detecting regional seismic phases. In Appendix 1, it is the most sensitive station, with the highest proportion of the REB events it has been associated with in the studied area since 2023. Therefore, PETK serves as a reference for other IMS station performance, and it is expected to be associated with all REB events within regional distances. However, there is one major obstacle to perfect performance: the high-amplitude noise, composed of signals that are coherent with those sought after J20 and J29.

Standard *SNR* estimates for the first P-phase arrivals in the REB associated with the immediate aftershocks of the J20 and J29 events are frequently below the automatic detection thresholds and are added by the IDC analysts. (An arrival is a detection with the estimated parameters such as arrival time, *SNRcc*, *SNR*, *CC*, amplitude, period, relative magnitude, station weight, etc. These parameters are used to create statistically significant hypotheses). The quality and reliability of these associated arrivals is lower relative to those obtained in the medium-to-low amplitude noise environment.

The *SNRcc* detection threshold was set in Figure 3 at two levels: 100 and 3.2. The higher value has to suppress all WCC arrivals, while the lower one is closer to optimal for finding valid arrivals when there are a large number of sought signals in the data. The relative efficiency of the REB events used as MEs varies for many reasons, but the most significant are related to the specificity of the corresponding signals, the hypocenter's location relative to the area of aftershock, and the depth dependent signal length. The MEs inside and outside the aftershock zone have to detect aftershocks and natural seismicity, respectively.

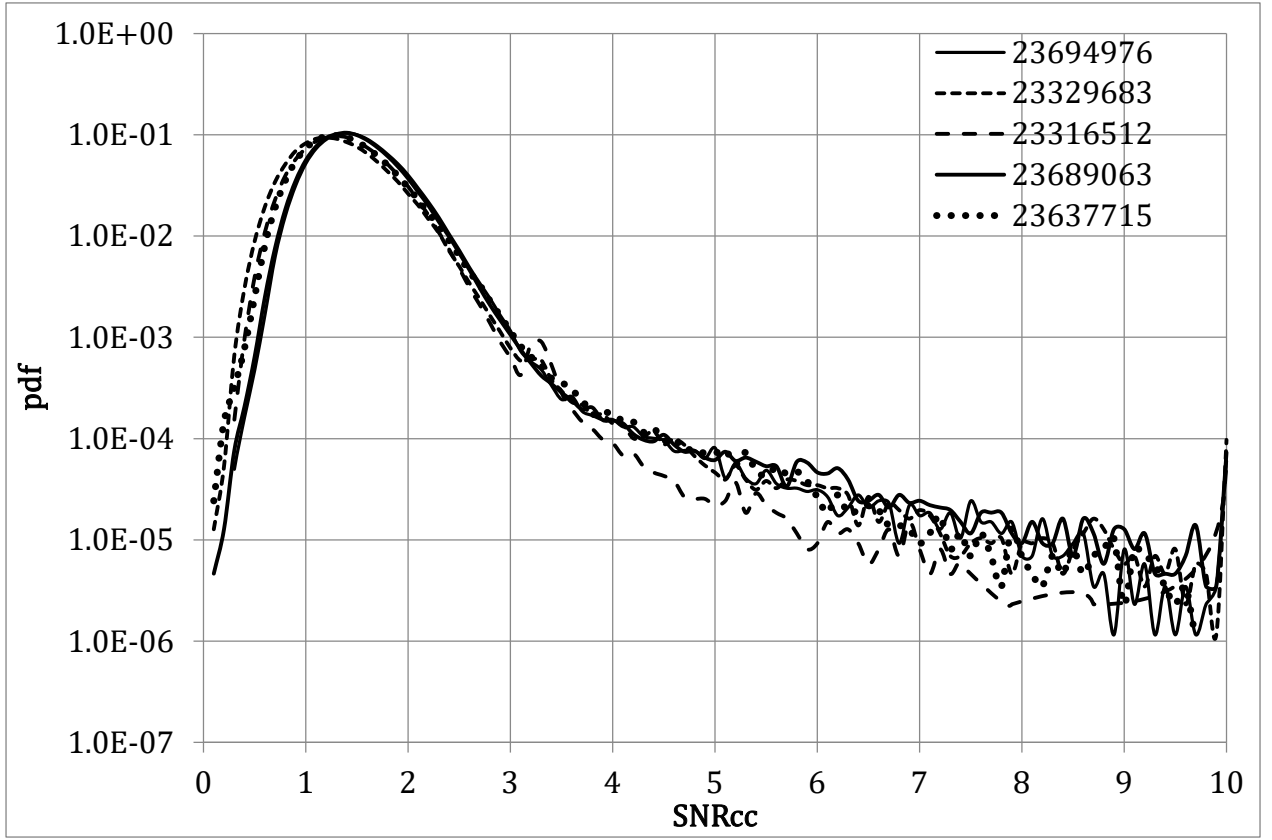

Figure 2. Examples of pdf for the first five MEs at station PDAR for July 25, 2025.

Remote MEs should not be able to detect aftershocks outside the search range confined by the virtual source grid in the phase association process. However, in some cases, the locations of event hypotheses may be poorly constrained due to a one-sided station distribution of stations associated with them. For example, if the azimuthal gap is large, say 300°, and there are only three or four defining stations with the associated P-phases, the difference in travel (origin) time residuals becomes less sensitive to the event location in the direction orthogonal to the direction to stations. As a result, there are more than 35,000 from approximately 800,0000 seismic events in the REB with the major semi-axis (Smaj) of the epicenter confidence ellipse larger than 500 km, and approximately 4000 events have Smaj larger than 2000 km. There are also few examples with Smaj>5000 km. These confidence ellipses indicate that the location is not reliable for many event hypotheses in the REB. From the stations with the highest weight, MKAR and KURK are almost in opposite direction to TXAR for the J20 and its aftershocks. Station PETK is located close to the J20 aftershock area, with the regional phases much more sensitive to distance, also in orthogonal to the event/station directions. Station PDYAR has a few templates and also at far-regional distances from the studied events. Therefore, the creation of false event hypotheses in XSEL is less probable than in the REB, where not all information of signal shapes is used [Schaff, Richards, 2025].

The PDF's for July 18th and July 29th are characterized by an *SNRcc* threshold of 100, illustrating the importance of the sought events in the data. On quiet days, the curves follow the expected random *SNRcc* distribution. However, on days with high aftershock activity, there are heavier tails above *SNRcc* > 2.2-2.5, with visible kinks around *SNRcc* ~3.0-3.3. Aftershocks of the J20 earthquake are visible in the pdf curves with peaks around *SNRcc*=3.3. A threshold of 3.2 indicates that the peak *SNRcc* value for arrival exceed the threshold. The efficiency of the two MEs differs, with the ME(orid=93694976) having a much higher curves above the threshold than compared to ME(orid=93689063). This difference should also be reflected in their respective inputs to the XSEL bulletin.

The event hypothesis with the highest total weight of all the associated stations wins out in potential conflicts for a shared phase between two or more MEs. This total weight is called the "event weight" and determines its statistical significance. If *SNRcc* for a given associated phase is high, the probability of this arrival being a random outcome is lower, as shown by the pdf in Figure 3. A reliable event hypothesis must include one or more high-quality signals (with large *SNRcc*) at stations with high weights (see Appendix 1). For example, the probability for signal detection at PETK (the most efficient station with a weight of 1.0) can be estimated for July 18 from the corresponding pdf's in Figure 3. The detection rate for the ME(orid=23694976) with *SNRcc*>5 is below 5/hour and for the ME(orid=23689063) there are no such detections at all.

The association of seismic arrivals with event hypotheses in the XSEL processing pipeline is based on a grid search around the set of MEs. For each master, a unique grid is tuned by size and node spacing to the slowness of the phases from the stations. The event hypotheses can be located only in the grid nodes and thus the phase association process is local to the given ME. We call this process Local Association (LA) to stress the contrast with Global Association used by the IDC and other seismological agencies. As a result of the LA applied to each and every ME, we obtain a final set of event hypotheses within a one day period, which minimizes the RMS origin time residuals for the associated arrivals.

An event hypothesis must have at least three first P-phase arrivals ($P_g$, $P_n$, P, PKP, $PKP_{ab}$, $PRK_{bc}$) associated, with only one allowed per station. The associated arrivals have to provide the sum of the weights of the corresponding stations above a predefined threshold. For each event hypothesis, there must be a high-quality arrival with a very low probability of being false, as defined by the *SNRcc*. For a grid node of the virtual sought event locations, the origin time for a virtual arrival is calculated using the ME/station travel time corrected for the grid node position relative to the ME and the theoretical/empirical slowness of the P-phase at each associated station. To be associated, an arrival has to have an origin time residual, i.e. the difference with the event origin time averaged over all associated stations, below the allowed tolerance. This tolerance varies from 2 s to 0.5 s (with selected cases tested using 0.25 s and 3 s). The lower the tolerance, the greater statistical significance of a given event hypothesis, as the probability of a random occurrence and their simultaneous association at two separate stations would decrease rapidly. We retain the average number of detections below 60 per hour, *i. e.* one every 60 seconds.

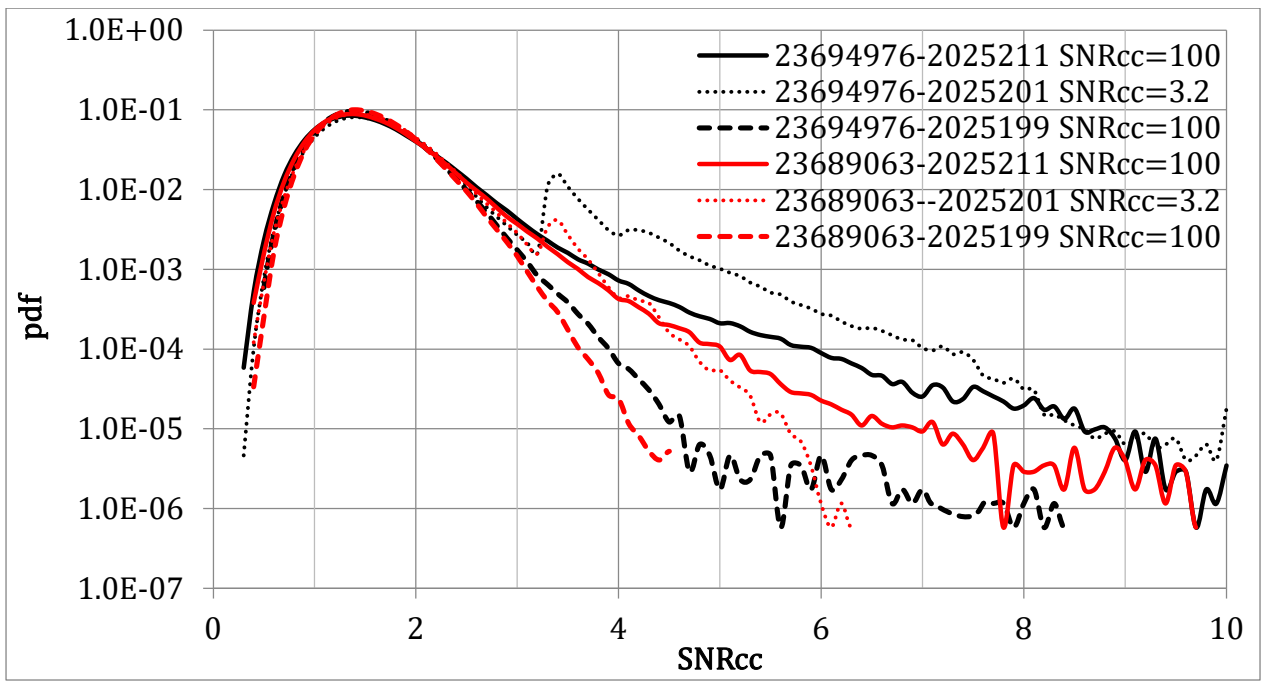

Figure 3. The pdf curves for *SNRcc* at station PETK on July 18 (2025199), July 20 (2025201), and July 30 (2025211), 2025. The detection thresholds for July 20 are 3.2 and 100 for July 18 and 30. Two MEs are presented.

The set of LA defining parameters has some other variables that are of lower importance. Trade-offs exist between different parameters, for example, lower *SNRcc* thresholds lead to an increased number of arrivals and a statistically larger average number of valid signals associated with a larger number of event hypotheses. For an event hypothesis *ceteris paribus*, the larger the number of associated arrivals, the higher its statistical significance. The lower the origin time tolerance, the more significant an event is for a given number of associated arrivals. An optimal set of thresholds and tolerances is hardly to be reached as the conditions for detection vary with time. Therefore, there is no need to determine an optimal set of parameters. The principal task is to find a set of reliable event hypotheses, with a part of them matching the REB and the rest of the set new to the REB. The epicenters of XSEL events that match REB events, with a depth fixed at the free surface, may be shifted by tens of kilometers or more.

**Results**

The number of events missed by the REB in the low magnitude range can be estimated by using recurrence curves. Figure 4 presents recurrence curves, as obtained from the REB events within the studied area in three depth ranges, and illustrates the deficit of REB events as a function of magnitude. The completeness threshold magnitude for a focal depth of 0 km is between 3.8 and 4.2. For the magnitude range from 3.6 to 3.8 there are ~3,000 to 4,000 missing events for the entire period of instrumental observations by the IDC. For the deepest events, the threshold magnitude is between 3.2 and 3.4, and hundreds of events are missing above the measured recurrence curve down to a magnitude of 2.0. The WCC-based methods are able to find many of these missing events. The REB consistency can also be improved as the number of associated phases, especially for smaller events, can be increased and the events physically close in space can be gathered around the ME which is the most relevant to them. This may reshuffle the REB and improve the absolute and relative location accuracy important for various seismic and mechanical research.

We are looking for potential events in the QZ before the J29 event. Figure 1 demonstrates that the J20 rupture failed to progress through this zone and a significant concentration of aftershocks from the REB is found just near its border. There are various reasons why the dynamic stress release in the J20 rupture was locked. One of the simplest explanations is that a mechanically strong asperity on the path of the J20 rupture stopped its progress. Since the asperity is intact and seemingly far from mechanical failure just after the J20, there should not be any large magnitude seismicity within its volume with the event lengths compatible to its size. However, there may be weaker seismicity activity with small-size events, which can be found by the detection methods with sensitivity higher than that of beamforming.

There are two sets of data required in order to confirm that the observed seismicity in the quiet zone is real. The ten days between the 20th (**jdate**=2025201) and 29th (2025210) of July have to be processed together with the days when no seismic events were found by the IDC in the studied area. These quiet days serve as a reference for WCC performance. We selected five

quiet days with **jdate**: 2024055 - 2024057, 2025093 and 2025141. Such days occur infrequently (~14% of the days between January 1, 2023 and July 20, 2025) in the area 45°N-65°N, 145°E-165°E, and the three-day gap in 2024 is an outstanding episode - there have only been two such gaps since January 2023 and four since January 2003. No longer gaps have occurred since the REB began in 2001. In the studied quiet zone itself, there were many 4-day and even longer (up to 30) periods without REB events.

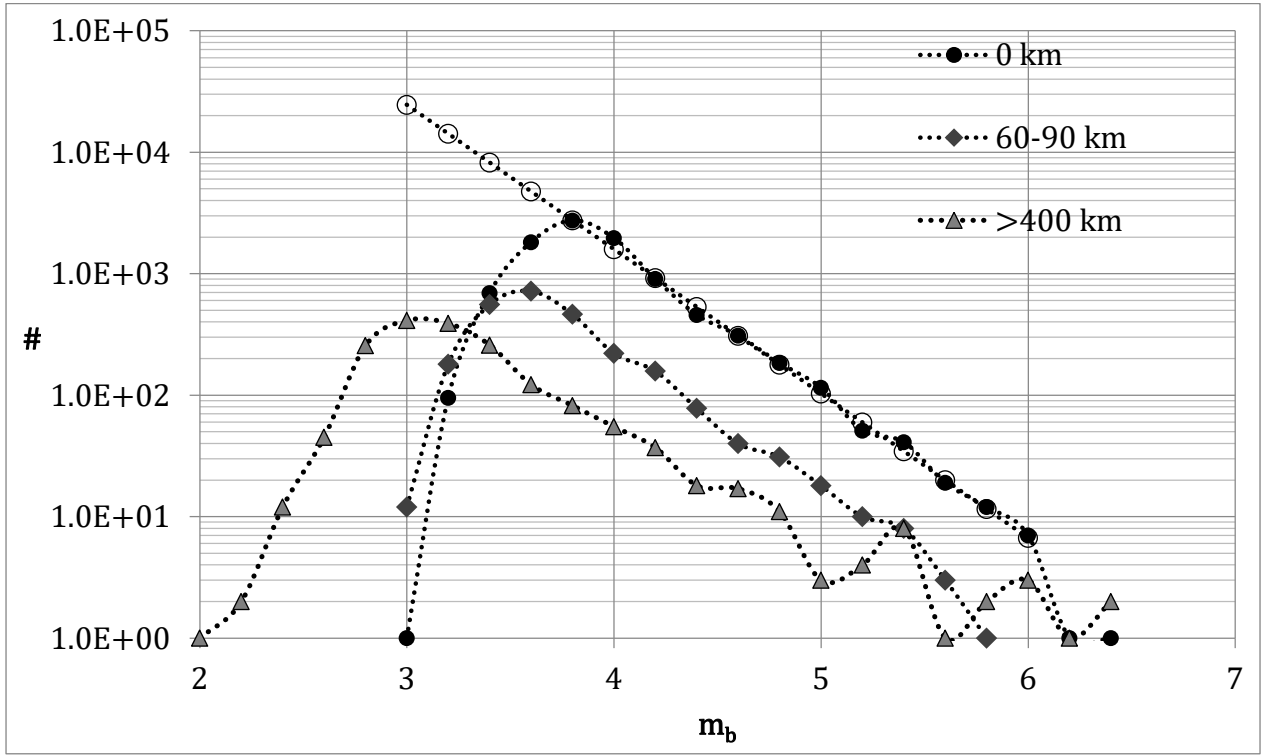

Figure 4. Depth dependent recurrence curves for the Kamchatka region as obtained from the REB since 2001. The extrapolated exponential trend line for depth=0 km illustrates the increasing deficit of found events with decreasing magnitude. The REB is complete to the threshold magnitude between 3.8 and 4.2.

The final product of the WCC processing, the XSEL, contains automatically generated event hypotheses, which can be divided into three groups according to the shared physical signals from the XSEL and REB. Groups 1 and 2 include the XSEL events with one or more REB arrival(s) matched. The match is defined by the arrival time difference within ±20 s (±10 s for each of the two compared arrivals) and is calculated automatically by a pair-wise comparison of XSEL and REB events at each station. Group 1 includes the XSEL events matching those REB events which were created from the SEL3 seed events, i.e. from the final product of IDC automatic processing. This group is called "SEL".

Group 2 contains XSEL events that match the REB events obtained by IDC analysts, not from SEL3 seeds, as a start point. It is called "REB". This division is essential for assessing the performance of the WCC relative to IDC automatic processing and for determining the reliability of XSEL event hypotheses. In the IDC, the events in the "REB" group are created through practically a random search process that is not based on continuous data analysis and/or review. This approach has clear limitations that can result in numerous missed REB-ready events matching the EDC [Bobrov *et al*., 2014].

The XSEL events not having matching detections with the events in the REB belong to the group "NEW". They have at least the same statistical significance as in the "REB" group, and, by design, are REB-ready [Bobrov et al., 2014]. We focus on finding events that were missed by the REB in the QZ. These "NEW" group of events likely represent the missed REB events and thus express the seismicity that was not detected below the IDC detection threshold.

The probability of an XSEL event hypothesis in the "NEW" group (or a new XSEL event) is defined by the parameters of the local association and conflict resolution that are numerous and tuned in the REB matching procedure. We have repeated the LA process in a wide range of defining parameters using the same arrivals lists. The most important parameters are:

1) The event weight is defined by the weights of associated stations, with the minimum allowed value from 1.8 to 2.5. The lower value corresponds to "weak" conditions allowing numerous combinations of 3 associated stations from the list in Appendix 1 to match the rule. The upper value requires the association of three from the top four stations. This upper value defines "strict" conditions.

2) The weight of at least one associated station has to be between 0.800 and 0.855 for an event with less than 7 associated stations. There are 5 station for the lower (weak) threshold and 3 for the upper (strict) threshold in Appendix 1.

3) The quality (probability), as defined by *SNRcc*, for at least one associated arrival has to be above 4.5 for weak and 5.5 for strict cases. For XSEL events with 3 or 4 associated stations, the *SNRcc* quality rule is set for the station with the weight above that in 2).

4) The size of the virtual location grid is 90 km in radius ($R_v$) (weak) and 48 km (strict). Valid event hypotheses have to be at distances less than $0.9 \cdot R_v$ from the corresponding ME. The hypotheses beyond this distance near the grid edge are rejected.

The origin time residual in the LA varies between 0.5 s and 2.0 s, with the latter value adapted by the REB events for the first P-phase. Many REB events would not be created if the 0.5 s origin time tolerance is applied. The WCC relative location accuracy is based on the precise arrival time estimates allowed by the form shape similarity between the template and sought signals. (Instructively, a 0.25 s tolerance, which is an exceptionally strict constraint, still allows to find a larger part of REB matching events in the XSEL.) The lower is the origin time tolerance, the higher is the event statistical significance since random arrivals have rapidly decreasing probability to be associated with an event hypothesis. As the XSEL is an automatic bulletin, the final decision on an event to be valid should belong to an experienced analyst. The overall experience gained from the interactive review at the IDC is invested in REB. Then, the XSEL events that do not match and those that match any REB event are equally likely to be valid, as they have similar statistical properties.

Table 1 presents the overall statistics for the WCC process over 5 days without REB events, and 10 days between July 20 and 29. Since the mainshock of the J29 event occurred at 23:24 (UTC), this day is processed between 0-22 hours. This should not affect the overall result. The evolution of XSEL over the studied period has two distinct components. The number of found (matched) REB and SEL3 events reflects the decrease in the occurrence rate in accordance with the Omori law. The number of new XSEL events depends on the level of seismic activity and the ambient noise level at relevant stations. Seismic activity is much higher at all magnitude levels after events as such J20 and J29. The number of new XSEL events after the J20 in Table 1 is much larger than that observed during the five quiet days. At the same time, the microseismic noise level after the J20 is elevated at all stations, and this noise is coherent with the signals generated by found and unfound aftershocks of the J20 earthquake. Higher noise prevents detection of low-amplitude signals from weak events. As the noise level decreases with the elapsed time, WCC regains the capability to find smaller and smaller signals to create XSEL hypotheses. The interaction between these two factors retains the number of new XSEL events at almost the same level for 10 days following the J20 event.

This is important to emphasize that the number of XSEL events matching the SEL3 and REB events can be larger than the number of the events in the IDC bulletins, since more than one XSEL event can match arrivals(s) from one REB/SEL3 event. There are a few reasons for this observation. Firstly, the REB events may have associated arrivals from different physical sources. These events are referred to as "joint" at the IDC, and are typically observed in aftershock sequences of the largest earthquakes. The XSEL may have event hypotheses created for these different sources, and thus one REB event may have matched phases with a few XSEL events. Secondly, as an automatic bulletin, the XSEL may create false hypotheses based on a mixture of primary and secondary phases of the physical events to be found. This is the case of interactive analysis as applied to the SEL3 automatic bulletin. We do not exclude the XSEL events matching the same REB, as they can be valid.

The REB match rate in Table 1 is lower during the days with the highest seismicity and increases to 100% when approaching July 29. The strict XSEL version matches a lower number of REB/SEL3 events, and the match rate decreases with decreasing origin time tolerance, Δt. This effect is obvious, but its extent is not typical, as the number of matches does not drop fast. For the weak parameter set on July 20 (2025201), the number of matched SEL3 events drops from 185 (from the total of 186 in the SEL3) for Δt=2s , to 180 for Δt=1s, and to 165 for Δt=0.5s. For the strict case at the same day, the match rate is 168, 167, and 165, respectively. This effect is observed on the background of a faster fall in the total number of XSEL events generate for these three Δt values: 592, 529, and 479. For lower seismic activity, the effect of Δt on the match rate is practically not observed. Therefore, the XSEL events in the "NEW" category (new XSEL events) are likely not created by random association, as the match rate does not change. Otherwise, the XSEL events with random associations would disappear randomly resulting in the match rate decrease. The new events obtained for larger Δt values are predominantly valid. At the end of the day, they need to have an interactive review in order to be promoted to the REB, as it is mandatory for the SEL3 events.

The number of new XSEL events and the total number of XSEL events are both important for the analysis of seismic regime within the QZ adjacent to the asperity. The XSEL hypotheses matching the REB events should be out of the QZ and asperity, maybe except for the "joint" REB events. The new XSEL events can occur within the QZ/asperity and Figure 5 demonstrates this observation showing their spatial distribution for days between July 20 and 29. The strict and weak XSEL versions illustrate the range of potentially missed REB events and their locations. In Figure 5, panels a) and b) present two origin time tolerances - 2 s and 0.5 s, respectively. In both versions, there are numerous event hypotheses created within the QZ/asperity volume, including the events with depth estimates. As the signals associated with the new XSEL events are very close to the ambient noise level, the relative magnitude estimates are rather biased as representing the noise amplitude. Appendix 3 lists the new XSEL events shown in Figure 5. There are several new parameters in Appendix 3 related to the WCC processing: the average and total |CC| values as well as the event weight (time 10 for internal processing reasons). The relative $m_b$ is obtained by averaging station relative magnitudes [Bobrov et al., 2014].

It is worth to compare the new XSEL events with those that match the SEL3 and REB events listed in Appendix 4. There are no statistically significant differences between the weakest REB matching and the average new XSEL events. The interactive review of the new XSEL events has a significant feature for the interactive analysis - low *SNR* values, often at the level of noise. Standard IDC automatic processing does rejects valid low-*SNR* signal as the number of false detections grows exponentially with the decreasing detection threshold. There is a specific measure in the WCC processing to retain *SNR* (following the IDC definition) at a signal visibility level defined by detection threshold *SNR* between 1.8 and 2.2 for IMS stations.

Table 1. Selected statistics of the WCC processing.

| Δt, s | Date | Start, h | End, h | REB | | | | Strict | | | | | | | Total XSEL | Weak | | | | | | | Total XSEL |
|---|---|---|---|---|---|---|---|---|---|---|---|---|---|---|---|---|---|---|---|---|---|---|---|
| | | | | SEL | REB | Total | SEL/Total | SEL | REB | Total REB | NEW | LA found | CR found | SEL found | | SEL | REB | Total REB | NEW | LA found | CR found | SEL found | |
| 2 | 2024055 | 0 | 23 | 0 | 0 | 0 | NA | 0 | 0 | 0 | 36 | 0 | 0 | 0 | 36 | 0 | 0 | 0 | 227 | 0 | 0 | 0 | 227 |
| 2 | 2024056 | 0 | 23 | 0 | 0 | 0 | NA | 0 | 0 | 0 | 24 | 0 | 0 | 0 | 24 | 0 | 0 | 0 | 194 | 0 | 0 | 0 | 194 |
| 2 | 2024057 | 0 | 23 | 0 | 0 | 0 | NA | 0 | 0 | 0 | 37 | 0 | 0 | 0 | 37 | 0 | 0 | 0 | 222 | 0 | 0 | 0 | 222 |
| 2 | 2025093 | 0 | 23 | 0 | 0 | 0 | NA | 0 | 0 | 0 | 27 | 0 | 0 | 0 | 27 | 0 | 0 | 0 | 211 | 0 | 0 | 0 | 211 |
| 2 | 2025141 | 0 | 23 | 0 | 0 | 0 | NA | 0 | 0 | 0 | 16 | 0 | 0 | 0 | 16 | 0 | 0 | 0 | 138 | 0 | 0 | 0 | 138 |
| | Average | | | | | | | | | | 28 | | | | | | | | 198.4 | | | | |
| | St.dev. | | | | | | | | | | 8.7 | | | | | | | | 36.1 | | | | |
| | Total | | | | | | | | | | | | | | **140** | | | | | | | | **992** |
| 2 | 2025201 | 6 | 23 | 186 | 49 | 235 | 0.79 | 203 | 16 | 219 | 42 | 208 | 192 | 168 | 261 | 296 | 38 | 334 | 258 | 229 | 223 | 185 | 592 |
| 2 | 2025202 | 0 | 23 | 114 | 29 | 143 | 0.80 | 131 | 21 | 152 | 46 | 130 | 127 | 108 | 198 | 176 | 27 | 203 | 321 | 135 | 134 | 110 | 524 |
| 2 | 2025203 | 0 | 23 | 83 | 34 | 117 | 0.71 | 107 | 18 | 125 | 42 | 104 | 99 | 78 | 167 | 145 | 34 | 179 | 291 | 113 | 112 | 83 | 470 |
| 2 | 2025204 | 0 | 23 | 26 | 3 | 29 | 0.90 | 33 | 2 | 35 | 80 | 28 | 27 | 25 | 115 | 39 | 3 | 42 | 306 | 28 | 28 | 26 | 348 |
| 2 | 2025205 | 0 | 23 | 33 | 12 | 45 | 0.73 | 40 | 10 | 50 | 67 | 43 | 43 | 32 | 117 | 56 | 11 | 67 | 314 | 45 | 45 | 33 | 381 |
| 2 | 2025206 | 0 | 23 | 23 | 5 | 28 | 0.82 | 33 | 5 | 38 | 87 | 28 | 28 | 23 | 125 | 44 | 4 | 48 | 349 | 28 | 27 | 23 | 397 |
| 2 | 2025207 | 0 | 23 | 18 | 2 | 20 | 0.90 | 22 | 2 | 24 | 61 | 20 | 20 | 18 | 85 | 37 | 3 | 40 | 250 | 19 | 19 | 18 | 290 |
| 2 | 2025208 | 0 | 23 | 6 | 3 | 9 | 0.67 | 6 | 1 | 7 | 54 | 9 | 8 | 6 | 61 | 8 | 2 | 10 | 248 | 9 | 9 | 6 | 258 |
| 2 | 2025209 | 0 | 23 | 5 | 2 | 7 | 0.71 | 5 | 1 | 6 | 66 | 6 | 6 | 5 | 72 | 9 | 1 | 10 | 306 | 7 | 6 | 5 | 316 |
| 2 | 2025210 | 0 | 22 | 9 | 0 | 9 | 1.00 | 9 | 0 | 9 | 61 | 8 | 8 | 8 | 70 | 12 | 0 | 12 | 337 | 8 | 8 | 8 | 349 |
| | Total | | | 503 | 139 | 642 | | 589 | 76 | 665 | 606 | 584 | 558 | 471 | **1271** | 822 | 123 | 945 | **2980** | 621 | 611 | 497 | 3925 |
| 1 | 2024055 | 0 | 23 | 0 | 0 | 0 | NA | 0 | 0 | 0 | 22 | 0 | 0 | 0 | 22 | 0 | 0 | 0 | 174 | 0 | 0 | 0 | 174 |
| 1 | 2024056 | 0 | 23 | 0 | 0 | 0 | NA | 0 | 0 | 0 | 20 | 0 | 0 | 0 | 20 | 0 | 0 | 0 | 133 | 0 | 0 | 0 | 133 |
| 1 | 2024057 | 0 | 23 | 0 | 0 | 0 | NA | 0 | 0 | 0 | 23 | 0 | 0 | 0 | 23 | 0 | 0 | 0 | 163 | 0 | 0 | 0 | 163 |
| 1 | 2025093 | 0 | 23 | 0 | 0 | 0 | NA | 0 | 0 | 0 | 14 | 0 | 0 | 0 | 14 | 0 | 0 | 0 | 155 | 0 | 0 | 0 | 155 |
| 1 | 2025141 | 0 | 23 | 0 | 0 | 0 | NA | 0 | 0 | 0 | 8 | 0 | 0 | 0 | 8 | 0 | 0 | 0 | 109 | 0 | 0 | 0 | 109 |
| | Average | | | | | | | | | | 17.4 | | | | | | | | 146.8 | | | | |
| | St.dev. | | | | | | | | | | 6.3 | | | | | | | | 25.9 | | | | |
| | Total | | | | | | | | | | | | | | **87** | | | | | | | | **734** |
| 1 | 2025201 | 6 | 23 | 186 | 49 | 235 | 0.79 | 217 | 13 | 230 | 24 | 208 | 189 | 167 | 254 | 290 | 32 | 322 | 207 | 225 | 215 | 180 | 529 |
| 1 | 2025202 | 0 | 23 | 114 | 29 | 143 | 0.80 | 134 | 16 | 150 | 27 | 127 | 125 | 108 | 177 | 183 | 29 | 212 | 252 | 133 | 131 | 108 | 464 |
| 1 | 2025203 | 0 | 23 | 83 | 34 | 117 | 0.71 | 112 | 21 | 133 | 27 | 101 | 98 | 78 | 160 | 143 | 38 | 181 | 211 | 113 | 112 | 83 | 392 |
| 1 | 2025204 | 0 | 23 | 26 | 3 | 29 | 0.90 | 29 | 2 | 31 | 50 | 27 | 27 | 25 | 81 | 36 | 2 | 38 | 230 | 28 | 28 | 26 | 268 |
| 1 | 2025205 | 0 | 23 | 33 | 12 | 45 | 0.73 | 39 | 9 | 48 | 47 | 42 | 42 | 31 | 95 | 52 | 13 | 65 | 239 | 45 | 45 | 33 | 304 |
| 1 | 2025206 | 0 | 23 | 23 | 5 | 28 | 0.82 | 31 | 4 | 35 | 61 | 27 | 27 | 23 | 96 | 40 | 4 | 44 | 266 | 28 | 27 | 23 | 310 |
| 1 | 2025207 | 0 | 23 | 18 | 2 | 20 | 0.90 | 19 | 1 | 20 | 42 | 19 | 19 | 18 | 62 | 34 | 1 | 35 | 175 | 19 | 19 | 18 | 210 |
| 1 | 2025208 | 0 | 23 | 6 | 3 | 9 | 0.67 | 8 | 1 | 9 | 36 | 8 | 8 | 6 | 45 | 8 | 2 | 10 | 175 | 9 | 9 | 6 | 185 |
| 1 | 2025209 | 0 | 23 | 5 | 2 | 7 | 0.71 | 8 | 0 | 8 | 43 | 5 | 5 | 5 | 51 | 10 | 2 | 12 | 225 | 7 | 7 | 5 | 237 |
| 1 | 2025210 | 0 | 22 | 9 | 0 | 9 | 1.00 | 9 | 0 | 9 | 40 | 7 | 7 | 7 | 49 | 12 | 0 | 12 | 260 | 8 | 8 | 8 | 272 |
| | Total | | | 503 | 139 | 642 | | 606 | 67 | 673 | **397** | 571 | 547 | 468 | **1070** | 808 | 123 | 931 | **2240** | 615 | 601 | 490 | 3171 |
| 0.5 | 2024055 | 0 | 23 | 0 | 0 | 0 | NA | 0 | 0 | 0 | 16 | 0 | 0 | 0 | 16 | 0 | 0 | 0 | 121 | 0 | 0 | 0 | 121 |
| 0.5 | 2024056 | 0 | 23 | 0 | 0 | 0 | NA | 0 | 0 | 0 | 11 | 0 | 0 | 0 | 11 | 0 | 0 | 0 | 92 | 0 | 0 | 0 | 92 |
| 0.5 | 2024057 | 0 | 23 | 0 | 0 | 0 | NA | 0 | 0 | 0 | 12 | 0 | 0 | 0 | 12 | 0 | 0 | 0 | 104 | 0 | 0 | 0 | 104 |
| 0.5 | 2025093 | 0 | 23 | 0 | 0 | 0 | NA | 0 | 0 | 0 | 8 | 0 | 0 | 0 | 8 | 0 | 0 | 0 | 103 | 0 | 0 | 0 | 103 |
| 0.5 | 2025141 | 0 | 23 | 0 | 0 | 0 | NA | 0 | 0 | 0 | 7 | 0 | 0 | 0 | 7 | 0 | 0 | 0 | 70 | 0 | 0 | 0 | 70 |
| | Average | | | | | | | | | | 10.8 | | | | | | | | 98 | | | | |
| | St.dev. | | | | | | | | | | 3.6 | | | | | | | | 18.8 | | | | |
| | Total | | | | | | | | | | | | | | **54** | | | | | | | | **490** |
| 0.5 | 2025201 | 6 | 23 | 186 | 49 | 235 | 0.79 | 208 | 14 | 222 | 18 | 202 | 187 | 165 | 240 | 310 | 25 | 335 | 144 | 202 | 187 | 165 | 479 |
| 0.5 | 2025202 | 0 | 23 | 114 | 29 | 143 | 0.80 | 138 | 16 | 154 | 18 | 122 | 119 | 104 | 172 | 190 | 23 | 213 | 162 | 132 | 131 | 109 | 375 |
| 0.5 | 2025203 | 0 | 23 | 83 | 34 | 117 | 0.71 | 102 | 13 | 115 | 15 | 92 | 88 | 78 | 130 | 150 | 34 | 184 | 161 | 112 | 111 | 83 | 345 |
| 0.5 | 2025204 | 0 | 23 | 26 | 3 | 29 | 0.90 | 29 | 2 | 31 | 29 | 27 | 27 | 25 | 60 | 44 | 2 | 46 | 184 | 28 | 28 | 26 | 230 |
| 0.5 | 2025205 | 0 | 23 | 33 | 12 | 45 | 0.73 | 40 | 8 | 48 | 30 | 41 | 41 | 31 | 78 | 49 | 13 | 62 | 187 | 45 | 44 | 31 | 249 |
| 0.5 | 2025206 | 0 | 23 | 23 | 5 | 28 | 0.82 | 29 | 3 | 32 | 30 | 26 | 26 | 23 | 62 | 41 | 5 | 46 | 197 | 28 | 27 | 23 | 243 |
| 0.5 | 2025207 | 0 | 23 | 18 | 2 | 20 | 0.90 | 24 | 0 | 24 | 21 | 18 | 18 | 18 | 45 | 33 | 1 | 34 | 126 | 19 | 19 | 18 | 160 |
| 0.5 | 2025208 | 0 | 23 | 6 | 3 | 9 | 0.67 | 6 | 2 | 8 | 23 | 8 | 8 | 6 | 31 | 9 | 2 | 11 | 125 | 99 | 9 | 6 | 136 |
| 0.5 | 2025209 | 0 | 23 | 5 | 2 | 7 | 0.71 | 6 | 0 | 6 | 28 | 5 | 5 | 5 | 34 | 10 | 2 | 12 | 147 | 7 | 7 | 5 | 159 |
| 0.5 | 2025210 | 0 | 22 | 9 | 0 | 9 | 1.00 | 9 | 0 | 9 | 20 | 7 | 7 | 7 | 29 | 15 | 0 | 15 | 182 | 8 | 8 | 8 | 197 |
| | Total | | | 503 | 139 | 642 | | 591 | 58 | 649 | **232** | 548 | 526 | 462 | **881** | 851 | 107 | 958 | **1615** | 680 | 571 | 474 | 2573 |

| | |
|---|---|
| Δt, s | - Origin time residual |
| Start, h | - Start hour of the studied interval |
| End, h | - End hour of the studied interval |
| LA found | - Number of REB events matched in the LA process before conflict resolution, which may reject some of the REB matching events in LA |
| CR found | - Number of REB events matched in the CR process. Same as the number of XSEL events matching the REB/SEL3 events |
| SEL found | - Number of SEL3 events matched in the CR process. |
| Strict | - Strict set of LA/CR parameters |
| Weak | - Weak set of LA/CR parameters |
| SEL | - REB events having SEL3 seeds. Also the number of XSEL events matching the SEL events |
| REB | - Pure REB events created from scratch. Also the number of XSEL events matching the pure REB events |
| NEW | - XSEL events not matching any REB event |

a)

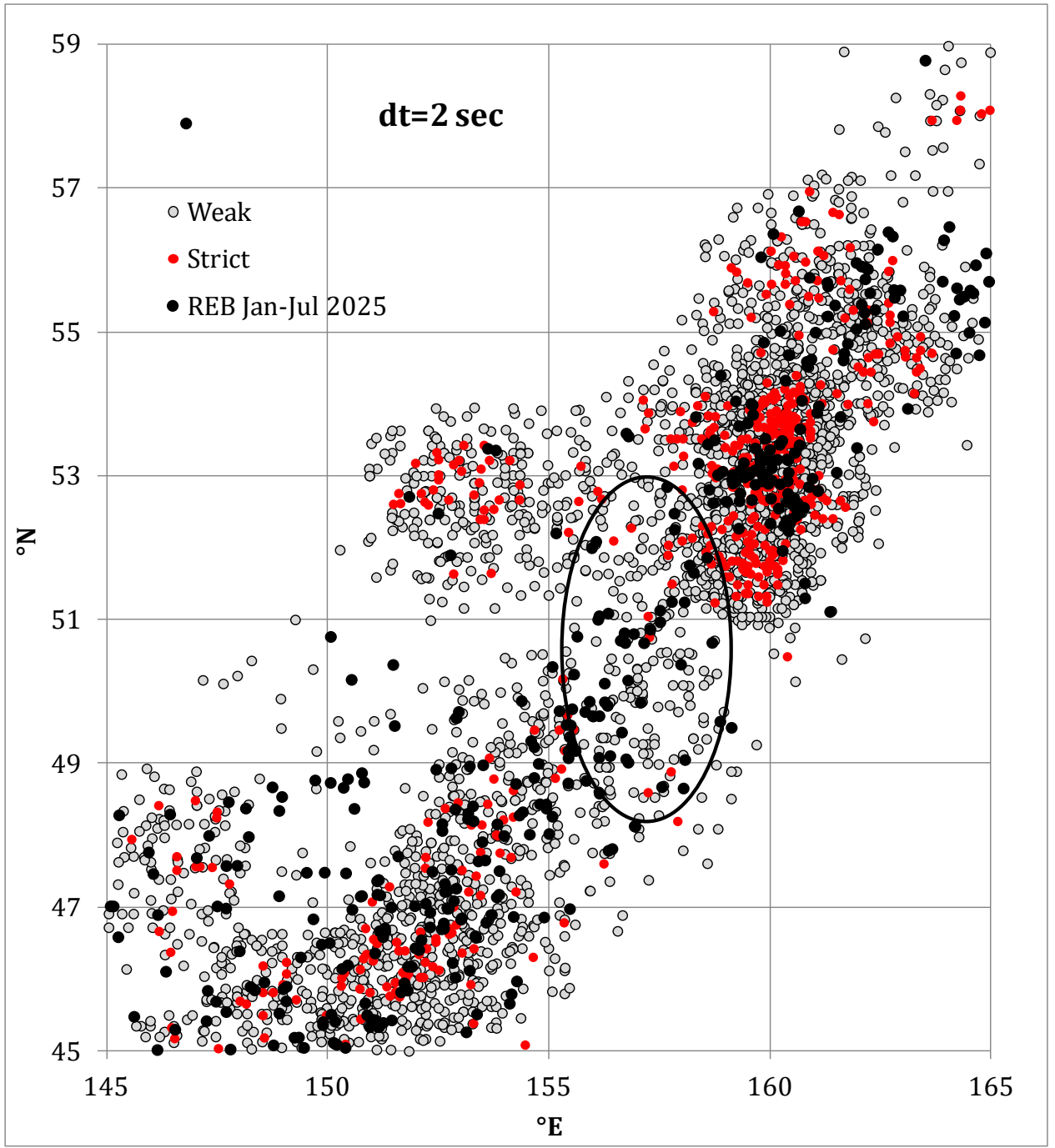

b)

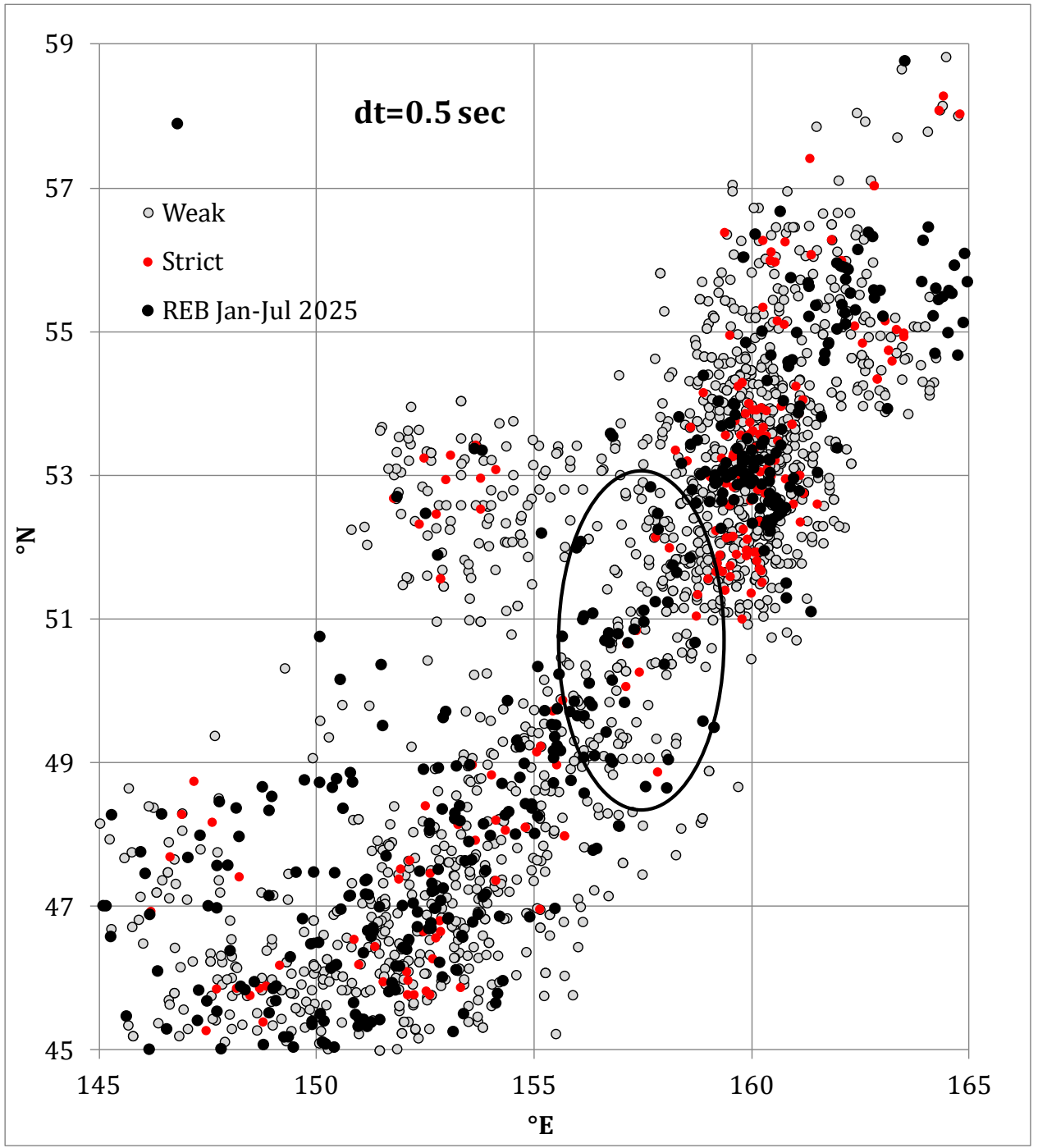

Figure 5. Distribution of the new XSEL events for the period from July 20 to July 29, 2025. Two versions of XSEL - strict and weak illustrate the range of events potentially missed in the REB. a) For origin time tolerance 2 s there are 2642 events in the weak XSEL version, and 483 - in the strict. b) Origin time tolerance 0.5 s: 1436 and 205 events in the weak and strict XSEL, respectively. Seismic events (390) from the REB for the period between January 1 and July 19 are shown for comparison.

Within the QZ, these new XSEL events are in agreement with the distribution density of the 390 REB events for the period between January 1 and July 19, 2025 (*e.g.,* Figure 5a). Many of them are concentrated in the epicenter zone of the J20 and J29. Some of them are aligned along the J29 rupture and parallel to it. The presence of REB events before July 20 in the studied quiet zone indicates that the asperity was able to generate observable seismicity, with two reliable REB events near the future rupture trace occurred on June, 21. Therefore, the absence of any REB events within the quiet zone during the days between July 20 and 29, accompanied by several hundreds of the J20 aftershocks on its border, is likely the result of the stress/strain changes likely induced by the J20. A majority of the new XSEL events are located in the aftershock zone of the J20 sequence and around, as their depths move them westwards along the subducting plate.

**Conclusion**

Seismic data processing based of the WCC method has a lower detection threshold for repeating signals. A set of well-located master events with precise source characteristics allows for extremely accurate relative location and phase association as a result of much more accurate estimates of arrival times, travel times and thus geographical coordinates. In seismically active regions like the Kamchatka Peninsula with the subducting Pacific plate, the events are so densely located that a WCC-based detector almost always has many potential master events to find very weak signals from adjacent events missed by global bulletins based on energy detectors. The smaller events one can find in a given region the better is the dynamic resolution of the pre-seismic, co-seismic, and post-seismic processes. The usage of WCC may provide an opportunity to obtain relevant seismic events related to the preparation of bigger earthquakes as the smallest events may manifest the reaction of the medium to increasing deformations and stresses [Schaff et al., 2025].

A still period in seismic activity after the July 20, 2025 M7.4 earthquake is observed along the rupture line of a much bigger and almost collocated earthquake occurred on July 29. It can be related to an asperity physically blocking propagation of the J20 rupture. As a result, there are no events in this quiet zone in the REB obtained from the IMS seismic data. The WCC-based automatic processing of the same IMS data was successful in creating of numerous event hypotheses in the automatic XSEL bulletin within the studied zone during the period between J20 and J29 events. These hypotheses have the same statistical significance as those matching the interactively reviewed REB events, both based and not based on SEL3 seed events. The new XSEL events in the quiet zone indicate the possibility of seismic activity in the asperity. The dynamic properties of such activity over the 10-day period may reveal some features related to the J29 preparation and mechanical destruction of the asperity.